\documentclass[aps,preprint,groupedaddress,showpacs]{revtex4}
\begin{document}

\title{Optical studies of gap, hopping energies and the
Anderson-Hubbard parameter in the zigzag-chain compound SrCuO$_2$}

\author{Z. V. Popovi\'c $^{a,c}$, V. A. Ivanov $^{a}$, M. J.
Konstantinovi\'c $^{b}$,  A. Cantarero$^{c}$, J. Mart\'{\i}nez-Pastor
$^{c}$,  D. Olgu\'{\i}n $^{b}$, M. I. Alonso $^{d}$, M. Garriga $^{d}$,
O. P. Khuong $^{a}$, A. Vietkin $^{e}$ and V. V. Moshchalkov
$^{a}$}

\affiliation{ $^a$ Laboratorium voor Vaste-Stoffysica en
Magnetisme, K. U. Leuven, Celestijnenlaan 200D, B-3001 Leuven,
Belgium }

\affiliation{$^b$ Max-Planck-Institut f\"{u}r Festk\"{o}rperforschung,
Heisenbergstrasse 1, D-70569 Stuttgart, F. R. Germany}

\affiliation{$^c$ Materials Science Institute, University of
Valencia, E46980 Paterna (Valencia), Spain}

\affiliation {$^d$ Institut de Ci\`{e}ncia de Materials de Barcelona,
CSIC, Campus de la UAB, 08193 Bellaterra, Spain}

\affiliation{$^e$ Department of Physics, Moscow State University,
119899 Moscow, Russia}

\begin{abstract}
We have investigated the electronic structure of the  zigzag
ladder (chain) compound SrCuO$_2$ combining polarized optical
absorption, reflection, photoreflectance and pseudo-dielectric
function measurements with the model calculations. These
measurements yield an energy gap of 1.42 eV (1.77 eV) at 300 K
along (perpendicular) to the Cu-O chains. We have found that the
lowest energy gap, the correlation gap, is temperature
independent. The electronic structure of this oxide is calculated
using both the local-spin-density-approximation with gradient
correction method, and the tight-binding theory for the correlated
electrons. The calculated density of electronic states for
non-correlated and correlated electrons shows
quasi-one-dimensional character. The correlation gap values of
1.42 eV  (indirect transition) and 1.88 eV (direct transition)
have been calculated with the electron hopping parameters $t$ =
0.30 eV (along a chain), $t_{yz}$ = 0.12 eV (between chains) and
the Anderson-Hubbard repulsion on copper sites U= 2.0 eV. We
concluded that SrCuO$_2$ belongs to the correlated-gap insulators.
\end{abstract}
\pacs{ 78.40.-q; 71.27.+a; 71.15.Ap; 71.20.Ps;}
\maketitle

\section{Introduction}
Strontium copper oxide, SrCuO$_2$, belongs to the new family of
quasi-one-dimensional (1D) insulators whose properties have been a
subject of intensive studies in recent years \cite {Hiroi, Kim,
Kim1, Motoyama, Matsuda, Misochko, Abrasev, Abrasev1, Popovic,
Nagasako, Filip, Wu}. This oxide, grown in single crystalline form
under ambient pressure, has an orthorhombic unit cell (space group
$D_{2h}^{17}$ - $Cmcm$) with parameters a=3.577 \AA, b=16.342 \AA,
c= 3.9182 \AA, Z=4 \cite {Taske, Matsushita}. SrCuO$_2$ has a
unique structure consisting of CuO$_4$ squares, mutually connected
via common edges, that form double copper zigzag chains (Fig. 1).

Magnetic susceptibility measurements of SrCuO$_2$ have revealed
that the Cu$^{2+}$ moments order antiferromagnetically below
$\sim$ 2 K \cite {Motoyama}. The exchange interaction energy $J$
is estimated to be $2100 \pm 200$ K \cite {Matsuda}. Very
recently, a static disorder spin structure (spin freezing) rather
than static three dimensional long range spin order is found in
SrCuO$_2$ using neutron spectroscopy \cite{Zaliznyak}. An
angle-resolved photoemission spectroscopy (ARPES) study of
SrCuO$_2$ shows the spin-charge separation in this oxide, as a
consequence of electron correlations \cite{Kim,Kim1}. The optical
phonons in SrCuO$_2$ have been investigated by measuring Raman
scattering \cite{Misochko,Abrasev,Abrasev1,Popovic} and
far-infrared reflectivity spectra on policrystalline
\cite{Abrasev} and single crystal \cite{Popovic} samples.

Electronic structure of SrCuO$_2$ has been calculated using the
linear-augmented-plane-wave (LAPW) method within the local-density
approximation (LDA) to the density functional theory
\cite{Nagasako}. As noted in Ref.\cite {Nagasako} for this type of
calculations the agreement with the ARPES data takes place only in
the region far from the Fermi level. Most recent band structure
calculations \cite {Filip, Wu} by the
local-spin-density-approximation (LSDA), including an on-site
Coulomb repulsion (LSDA+U method) have produced an insulating gap
of 1.63 eV for U=12.3 eV \cite{Filip} and 2.36 eV for U=5 eV \cite
{Wu}. Due to the lack of experimental data for the energy gap of
SrCuO$_2$, it was not possible to compare these data with
experiments. Quite recently, Rosner {\it et al.} \cite {Rosner}
calculated the electronic structure of SrCuO$_2$ using the LSDA
method, also. They found that the gap in the antiferromagnetic
phase of SrCuO$_2$, reported in previous publications \cite
{Filip,Wu}, is probably artificial and appears due to too small
number of the {\bf k}-points in the Brillouine zone taken in the
computation procedure.

In this paper we have applied complementary optical spectroscopy
techniques to study the electronic structure of SrCuO$_2$. We have
obtained an energy gap of 1.42 eV (1.77 eV) at 300 K along
(perpendicular to) the Cu-O chains. These values are compared with
the results of the electronic structure calculations completed by
us. We have established that the tight-binding method for the
correlated electrons, with  the hopping parameters $t$ = 0.30 eV
and $t_{yz}$ = 0.12 eV along legs and between legs, respectively,
(see Fig. 1) and the Anderson-Hubbard parameter $U$ = 2.0 eV
yields the energy gap values in an agreement with the experimental
results. We also have determined the density of electronic states
for non-correlated and correlated electrons, which show the
one-dimensional nature. According to the electronic structure
calculations and the experimental findings, we concluded that
SrCuO$_2$ is a correlated-gap insulator.

The remainder of this paper is organized as follows. Experimental
details are given in Sec. II. The results of electronic structure
calculations for non-correlated electrons (using the LSDA with
gradient corrections to density functional theory) are described
in Sec. III.A. The reason why we use this method is to obtain the
data necessary for the determination of the correlated electronic
structure and to compare our results with the previously published
ones \cite{Nagasako,Filip,Wu, Rosner}. In Sec. III.B, we applied
the tight-binding theory for correlated electrons taking into
account the realistic crystalline structure of SrCuO$_2$. We got
analytically the energy dispersion relations, the density of
states of correlated and non correlated electrons as well as the
correlation gap. Experimental measurements of dielectric function,
reflectivity, photoreflectance and optical transmission spectra
are given in Sec. IV. Section V contains the discussion of
experimental and theoretical results and the conclusions.

\section{Experimental details}

The present work was performed on a single crystal sample with a
size of about 15, 2 and 6 mm along the {\bf a}, {\bf b} and {\bf
c} axes, respectively. We used several optical spectroscopy
techniques. The pseudo-dielectric function was measured with the
help of a rotating-polarizer (analyzer) ellipsometer. A Xe-lamp
was used as a light source, a double monochromator with 1200
lines/mm gratings and an S20 photomultiplier tube as a detector.
The polarizer and analyzer were Rochon prisms.  The measurements
were carried out in the 1.2-5.6 eV energy range. For the energies
below 1.6 eV we used a halogen lamp as a light source and Si-photo
diode as a detector. Optical reflectivity and transmission spectra
were measured at room and liquid helium temperature in the
200-2500 nm spectral range with Perkin - Elmer Lambda 19
spectrophotometer. Photoreflectance measurements were performed at
room temperature under incident polarized light parallel to the
{\bf a}- and the {\bf c}-axes. The reflected light was dispersed
through a single 1/2 m monochromator and sinchronously detected by
Si-photodiode.

\section{Electronic structure calculations}

\subsection{LSDA with gradient corrections}

In the CuO$_2$-layered oxides, the electron dispersion branches at
the Fermi level are derived from the 3$d_{x^2-y^2}$ copper
orbitals, mixed with the corresponding O ${2p}$ orbitals. Other
occupied $3d$ Cu orbitals ($d_{3z^2-r^2}, d_{yz}, d_{xz}, d_{xy}$)
are located below the Fermi level and they are strongly hybridized
with oxygen bands. These cuprates are known as charge-transfer
insulating oxides. Contrary to the CuO$_2$-layer cuprates
(tetragonal crystal structure oxides), SrCuO$_2$ has an
orthorhombic crystal structure with the Cu-O zigzag chains and,
consequently, with the different electronic structure, as it will
be shown later.

For the ab-initio calculation of the electronic structure of
SrCuO$_2$ we used WIEN97 software package \cite {wien}. The
program allows us to compute the electronic structure within the
density functional theory by applying the LAPW method with a
simplified version of the generalized gradient approximation
\cite{Perdew1} for the exchange-correlation functional in the LSDA
description. This approximation takes into account the charge and
spin inhomogeneity in a material by including gradient corrections
in the energy functional and gives more precise ground state
energy \cite {Perdew} but does not improve the quasiparticle
spectra \cite {Dufek}.

The Figure 2 shows the calculated electron energy dispersions
along the high-symmetry lines of the Brillouin zone (BZ). The most
interesting feature of these calculations is a relatively large
dispersion along the $k_z$ direction parallel to the Cu-O chains
and dispersionless bands perpendicular to them, {\it i.e.} along
the $k_x$ and $k_y$ directions. Along the $\Gamma$-Z direction of
the BZ, the two bands, which cross the Fermi level, split
slightly, indicating that the interaction between the two
neighboring chains in the ladder is small, but still exists. The
total density of electronic states of SrCuO$_2$ is given in Fig. 3
in the energy range between -7 eV and 10 eV. The electron density
of states is calculated with 180 {\bf k} points in the irreducible
part of the BZ and it is in an agreement with previous
publications \cite{Nagasako,Rosner}. Because the main properties
depend on the electrons at the Fermi level, we paid a special
attention to the energy dispersions near $E_F$. In the inset of
Fig. 3 we show the partial density of electronic states in the
energy range from -1.5 eV to 1 eV. Figure 4 gives the contribution
from the different Cu $3d$ and O$2p$ orbitals to the partial
density of states in the energy range close to the Fermi level.
From the results given in Figs. 2-4 we concluded that:

{\it (i)} The electronic structure of SrCuO$_2$ is a
quasi-one-dimensional one.

{\it (ii)} The main contribution to the density of states in
SrCuO$_2$ near the Fermi level comes from the Cu $3d_{y^2-z^2}$
orbitals.

{\it (iii)} There is a small hybridization between the Cu
$3d_{y^2-z^2}$ and the O $2p_z, 2p_y$ orbitals.

These results are in an agreement with the X-ray absorption
spectroscopy data \cite {Knupfer}, which show that the holes in
SrCuO$_2$ have predominantly the $3d_{y^2-z^2}$ character (which
is analogous to the Cu $3d_{x^2-y^2}$ orbitals in notation for the
2D high-T$_c$ cuprates). The hole occupancy of $3d_{3x^2-r^2}$ is
less than 5\% \cite{Knupfer}.

However, the metallic state obtained in the framework of the LSDA
calculation (in Figs.2-4 the Fermi level $E_F$=0 is inside the
occupied band) does not allow us to compare these results with our
experiments.

The applied LAPW method with the gradient LSDA corrections (or
other LDA versions to the density functional) employs the
orbitally independent exchange-correlation potential, which can
not recognize different $d$-orbitals in the copper ion with an
open $d$-atomic shell, $Cu^{1+n}(3d^{10-n})$, of the SrCuO$_2$
material. As a result it gives a satisfactory band structure, but
can not overcome ''an energy gap problem''\cite{Fulde}. In
general, the local approximations to the density functional theory
(for a review see \cite{Fulde} and Refs. \cite {Gunnarsson,
Perdew2, Zein} for an inclusion of the spin-dependent
exchange-correlation potentials) are successful to describe the
ground state energies and the quantities, weakly depending on a
charge or spin density \cite{Yu, Hybertsen}, but the excited
states are beyond the scale of the density functional theory.

To alleviate an energy gap problem of L(S)DA to density functional
theory the LDA+U method has been proposed (for details see Ref.
\cite{Anisimov} and its generalization, LSDA+U \cite{Dudarev}).
These methods have extended LDA-like functionals with an addition
to them the on-site Hubbard terms and they are based on the
unrestricted Hartree-Fock wave-functions. For the Mott-Hubbard
insulating compounds with transition or rare-earth metals the
orbital-dependent potential $U$ leads to splitting of the
partially filled 3$d$- or 4$f$-bands forming the upper and the
lower Hubbard-like bands. The LDA+U functional gives more correct
antiferromagnetic properties than the LSDA one. Also LDA+U's are
relevant for study of the temperature induced phase transition
from an non-magnetic insulator to a metal with a local magnetic
moments and for study of the charge orders, where LSDA fails.
However, the used
mean-field correction for a single orbital, $U\widehat{n}_{\uparrow }%
\widehat{n}_{\downarrow }\longrightarrow $ $U\widehat{n}_{\uparrow
}n_{\downarrow }$($n_{\downarrow }$ is an expectation value for
the number of electrons with a spin projection $\downarrow $\
occupying a fixed partially filled orbital on a particular lattice
site), to an exchange-correlation potential reproduces only
qualitatively the correct physics for the Mott-Hubbard insulators.
Inclusion of the inter-orbital intra-atomic interactions (Coulomb
repulsion and Hund exchange) and often the antiferromagnetic
ordering \cite{Anisimov1} (regardless the absence of an
antiferromagnetism in material under consideration; {\it e.g},
SrCuO$_2$ is in paramagnetic state above 1.4K) does not change
significantly the qualitative character of the LDA+U results. The
obtained gap values differ by about 1eV from the experimental ones
\cite{Anisimov, Dudarev}. Also in the framework of LDA+U\ methods
the band structure is unsatisfactory and these methods produce
unphysical insulating states for transition metals since the
LDA+U's split even the partially-filled bands due to the
self-interaction corrections both for localized and extended
electron states. It is also known that the electron correlations
produce narrowing of the dispersion branches and consequently an
increase of the effective masses \cite{Olson, Manske, Merino}
which is outside the LDA+U domain.

As applied to SrCuO$_2$, the band gap values obtained in the
LDA+U\ calculations \cite{Filip, Wu} differ significantly from our
experimental findings (see Sec. IV). Besides that, a shift of the
energy bands due to a finite $U$ can not guarantee the correct
band extrema positions to identify the direct or indirect
character of the interband electronic transitions. In
Ref.\cite{Filip} the reported band gap 1.63 eV (the most close to
our measured value) corresponds to an unrealistically large
Anderson-Hubbard parameter $U$=12.3 eV.

Having all above mentioned in mind, we developed a model which
{\it introduces the electron correlations already in the zeroth
order of the applied perturbation theory with respect to the
hopping energies in a realistic lattice structure of SrCuO$_2$}
(Sec.III.B).

\subsection{The tight binding method for the correlated electrons}

According to our LSDA data (Figs.2-4) and also Ref.
\cite{Knupfer}, the main contribution to the electron density of
states close to the Fermi level comes from the $d_{y^2-z^2}$
orbitals, slightly hybridized with the O $2p$ orbitals. In
addition, our absorption measurements (see, Sec. IV) show that the
lowest energy gap is practically temperature independent (a charge
transfer gap depends on temperature in accordance with the
variation of the Cu-O distance with temperature). All these
important observations have lead us to the conclusion that the
band gap in SrCuO$_2$ is a consequence of strong electron
correlations, favoring the Mott-Hubbard like insulating state.

Although the density functional theory reduces the Schr\"{o}dinger
equation for electrons and atomic nuclei in solids to non-linear
single-particle equations (equations with self-consistent
potential, which itself depends on the solutions of the
equations), the explicit dependence of the ground state energy on
the ground state charge and spin density is unknown and such an
explicit functional may even not exist. As the starting many body
linear Schr\"{o}dinger equation, the problem of the non-linear
single-particle equations with the exchange-correlation potentials
is a formidable problem which can not be solved without additional
approximations. The exchange and correlations are often
approximated by the L(S)DA according to which the charge (spin)
density in the exchange-correlation potential of an electron gas
or a jellium model is replaced by the local density in a real
material. The strategy of the LDA approach is close to the Landau
fermi-liquid theory, where the energy is a functional with respect
to a single-particle partition function. But contrary to the
Landau theory, the density functional theory can calculate only
the ground state energy, distributions of charge and spin
densities and the quantities connected with them. The main problem
in our case is that LDA and its versions ({\it e.g.}, the one used
by us) do not give a clear knowledge about the Anderson-Hubbard
parameter (an essential on-site repulsion of the correlated
electrons) in narrow-energy band materials such as SrCuO$_2$.

The conventional tight-binding method $\left( U=0\right) $ for an
electronic structure in solids is self-contradictory. According to
it the eigen-functions of the Schr\"{o}dinger equation are
approximated using the electron wave-functions of the isolated
atoms, $i.e.$ a conventional tight-binding method is the more
suitable the greater interatomic distances in the crystal are. But
in this case the prevailing terms in the Hamiltonian, the strong
electron-electron interactions $U$, become even larger; they can
not be reduced to any mean field version and then the problem is
fully outside the domain of the standard Slater-Koster scheme. The
present tight-binding method for correlated electrons is based on
the hypothesis that the narrow-energy band material properties,
$e.g.$, SrCuO$_2$ in our case, are governed by the intraatomic
electron correlations $U$, exceeding considerably the transfer
energies.

A reasonable simulation of the many-body effects leads to
tremendous problems in terms of the conventional Fermi- or
Bose-operators permutation relations which do not encompass all
possibilities of the second quantization formalism. The
permutation relations for the Okubo-Hubbard X-operators are linear
with respect to theirselves operators. The necessity to introduce
the operators with more complicated permutation relations than the
fermionic and bosonic ones for study of the electron correlations
was indicated by Bogoljubov already in 1949 \cite{Bogoljubov}. We
consider the tunneling part of any correlated Hamiltonian as a
perturbation with respect to the strong electron correlations
included in the unperturbed part of the Hamiltonian. The
Hamiltonians with correlated electrons are rewritten in terms of
the basis and only basis vectors of the corresponding
superalgebra. To treatment of such Hamiltonians in the
$X$-operator framework is based on a rigorous successive method
which we followed in our previous publications \cite{Ivanov2,
Popovic1, Ivanov} and it will be also applied here. The systematic
perturbation theory is based on the generalized Wick's theorem as
an iteration procedure reducing the time-ordered product of $n$ of
X-operators to product of $n-\mathit{1}$ of thereof. The first
order self-energy is the tunneling matrix itself from the
perturbative Hamiltonian.

Here in the framework of the $su\left( 2,2\right) $ superalgebra
approach for the SrCuO$_2$ system we will neglect the scattering
of the correlated copper electrons by the spin and charge
fluctuations, aiming at comparing the electron spectra with the
conventional tight-binding calculations, which are done in the
first order of the transfer energy. Contrary to the standard
Hubbard model for the $s$-electrons with a single site per a unit
cell, the starting Hamiltonian includes a realistic unit cell (a
few sites) and the correlated electrons with the non-zero angular
momenta. In the framework of {\it the presented tight-binding
method for the correlated electrons} the non-spherical
wave-functions provide anisotropy of the hopping integrals in the
lattice. In the considered order of the perturbation theory we
will concentrate on the influence of the band structure effects
which are of significance for multicomponent systems such as
SrCuO$_2$.

The LSDA calculation of the electronic structure of SrCuO$_2$,
Figs.2-4, revealed that electron energy dispersions are governed
mainly by electrons in the zigzag ladder. The interladder coupling
is negligible because of the large (6.75 \AA) Cu atom distance
between neighboring ladders. A unit cell for the correlated
electron structure calculation includes two Cu-ions only, as
indicated by a dashed rectangular in Fig. 1(a). We assume that a
ladder unit Cu$^{1+n}$O$_2^{2-}$ has the total charge $-2$, {\it
i.e.} there is one hole, $n=1$, per a copper ion in the ladder. In
our minimal model, we neglected hybridization with the $p_y,
p_z-$orbitals of the intermediate oxygen atoms, assuming the
$d_{y^2-z^2}$ character of holes. Such approximation is common for
the 1D cuprates \cite{Tsutsui} because the Anderson-Hubbard
repulsion usually opens a gap between the $3d$ bands. Inside the
ladder, the arrangement of the Cu-atoms is such that the
directions from the Cu-ion in one leg to the two nearest Cu-ions
of the neighboring leg are almost at a right angle $\left( \sim
90^{\circ }\right) $, see Fig.1. In our consideration of the
electronic structure we assume that these directions form an ideal
right angle and the electron energy dispersions are governed
mainly by the correlated electrons in the single zigzag ladder.

Aiming at the determination of the Anderson-Hubbard parameter $U$
from the optically measured correlation gap values, the further
theoretical study is based on the realistic Hamiltonian with one
copper rung/dimer, $a-b$, per a unit cell:
\begin{eqnarray}
H &=&-2t\sum\limits_{p,\sigma }\cos p_z\left[ a_\sigma ^{+}\left(
p\right) a_\sigma ^{}\left( p\right) +b_\sigma ^{+}\left( p\right)
b_\sigma ^{}\left( p\right) \right] \nonumber
\\ &&-t_{yz}\sum\limits_{p,\sigma }\left( 1+e^{-ip_z}\right)
\left[ a_\sigma ^{+}\left( p\right) b_\sigma ^{}\left( p\right)
+H.c.\right] +U\sum\limits_{i=a,b}n_{\uparrow i}^jn_{\downarrow
i}^j-\mu \sum\limits_{i=a,b}n_i^j,
\end{eqnarray}
where $a,b$ denote chains (legs in the ladder directed along the
c-axis), $t$ is an amplitude of the carrier hopping along the
legs, $t_{yz}$ is a diagonal hopping amplitude between the legs,
$U$ is the Anderson-Hubbard repulsion of the
$d_{y^2-z^2}$-electrons on copper site and $\mu$ is the chemical
potential. In LDA band structure calculations the
electron-electron interactions are approximated by an effective
single-electron problem, whereas in our approach the
Anderson-Hubbard parameter $U$ is included explicitely in the
non-perturbative Green's function. Applying the $X-$operator
machinery \cite {Ivanov}, the correlated energy bands are governed
by zeros of the inverse Green's function for the $X$-operators in
the first order with respect to the tunneling matrix:
\begin{eqnarray}
\widehat{D}_p^{-1}\left( \omega \right)  &=&
\begin{tabular}{l}
$a$ $\{
\begin{tabular}{l}
0+ \\ -2
\end{tabular}
$ \\ $b$ $\{
\begin{tabular}{l}
0+ \\ -2
\end{tabular}
$
\end{tabular}
\left(
\begin{array}{llll}
\frac{-i\omega _n-\mu }{f_{0+}}+r & r & v & v \\ r &
\frac{-i\omega _n-\mu +U}{f_{-2}}+r & v & v \\ v^{\star} &
v^{\star} & \frac{-i\omega _n-\mu }{f_{0+}}+r & r \\ v^{\star} &
v^{\star} & r & \frac{-i\omega _n-\mu +U}{f_{-2}}+r
\end{array}
\right) ,
\end{eqnarray}
\\ where $r =-2t\cos
p_z$, and $v=-t_{yz}\left( 1+e^{-ip_z}\right)$. Here the
correlation factors, $f_{0+(-2)}$ , in the diagonal Green's
functions are determined by the fermion occupation $n$ per site.
For a considered nonmagnetic phase of SrCuO$_2$ they are
$f_{0+}=1-n/2$, $f_{-2}=n/2$ and all equal to $1/2 $ ($n=1$).
After an analytical continuation, $i\omega _n\rightarrow \xi
+i\delta $, from the secular equation $\left|
\widehat{D}_p^{-1}\left( \omega \right) \right| =0$ one can find
the four branches of the correlation energy dispersions in an
explicit form:

\begin{equation}
\xi _B^{\pm }\left( p\right) =\varepsilon _p^{1,2}+\sqrt{\left(
\varepsilon _p^{1,2}\right) ^2+\left( \frac U2\right) ^2}-\mu,
\end{equation}
\begin{eqnarray}
\xi _A^{\pm }\left( p\right) &=&\varepsilon _p^{1,2}-\sqrt{\left(
\varepsilon _p^{1,2}\right) ^2+\left( \frac U2\right) ^2}-\mu,
\end{eqnarray}
where $\varepsilon _p^{1,2} =-t\cos p_z\pm t_{yz}\cos
\frac{p_z}2$. For the derivation of these energy dispersions from
the four-by-four fold secular equation (see Eq.(2)) it was useful
to apply the proved theorem about the decomposition of the
determinant with respect to diagonal elements (see Appendix A in
Ref. \cite{Ivanov1}).

The lower correlated subbands, $\xi_A^{+}$ and $\xi_A^{-}$, are
completely occupied by the two holes from a unit cell of the
zigzag ladder. It is essential that the chemical potential $\mu$
should be found self-consistently from the equation for a particle
density $$n=T\sum\limits_{p,\omega,\alpha,\beta}e^{i\omega\delta}
D_{\alpha,\beta}(p,\omega_n),$$ where $\alpha$ and $\beta$ label
the rows and columns of the Green's function elements (c.f.
Eq.(2)). For $n=1$ (the case of SrCuO$_2$) the calculated chemical
potential is positioned in the middle of the energy gap ($\mu$=0).
For other densities, $n\neq1$, the used procedure generates a
correlated metal. The upper correlated
subbands, $\xi _B^{\pm }$, are empty for SrCuO$_2$. The nearest unoccupied energy band is $%
\xi _B^{-}$ and the correlation gap (indirect transition, see
Fig.5) in an electronic structure can be estimated as
\begin{eqnarray}
\Delta _0 &=&\min \xi _B^{-}\left( p\right) -\max \xi _A^{+}\left(
p\right) = \nonumber \\ &&\sqrt{\left( t+t_{yz}\right) ^2+\left(
\frac U2\right) ^2}+\sqrt{\left( t+\frac{t_{yz}^2}{8t}\right)
^2+\left( \frac U2\right) ^2}-\left(
2t+t_{yz}+\frac{t_{yz}^2}{8t}\right) .
\end{eqnarray}

For the non-correlated energies, $\varepsilon ^{1,2}=2\varepsilon
_p^{1,2}$ ({\it c.f.}, Eqs.(3,4)), the electron density of states
per atom, $\rho_0 $($\varepsilon $)$=\sum\limits_{p_z}$[$\delta $($%
\varepsilon -\varepsilon ^{\left( 1\right) }$)$+\delta
$($\varepsilon -\varepsilon ^{\left( 2\right) }$)], is defined
analytically as follows

\begin{equation}
\rho _0^{}\left( \varepsilon \right) =\frac 1{\pi
\sqrt{t_{yz}^2-4t\left( \varepsilon -2t\right)
}}\sum\limits_{i=1}^4\frac 1{\sqrt{1-x_i^2}},
\end{equation}
where
\begin{equation}
x_{1,2} =  \frac 1{4t}\left( -t_{yz}\pm \sqrt{t_{yz}^2-4t\left(
\varepsilon -2t\right) }\right) =-x_{4,3}.
\end{equation}

The electron-electron repulsion, $U$, splits the density of the
non-correlated electron states $\rho _0(\varepsilon)$. After the
transformation to the ''correlated'' variables, Eqs. (3) and (4),
one can get the correlated electron density of states as
\begin{eqnarray}
\rho \left( \xi \right)  &=&\frac{\xi ^2+\left( \frac U2\right) ^2}{\xi ^2}%
\rho _0\left( \frac{\xi ^2-\left( \frac U2\right) ^2}{\xi
^{}}\right) = \nonumber \\ &&\frac{\xi ^2+\left( \frac U2\right)
^2}{\xi ^2}\sum\limits_{p,\alpha =A,B}\frac{\xi _\alpha ^2\left(
p\right) }{\xi _\alpha ^2\left( p\right) +U^2}\delta (\xi
-\frac{\xi _\alpha \left( p\right) }2).   \\ && \nonumber
\end{eqnarray}

Figure 5(a) shows the correlated electron energy dispersions
(Eqs.(3) and (4)) along the Z and the Y symmetry directions of the
BZ. The correlated electron density of states is given in  Fig.
5(b). The overlap of the energy ranges for the energy dispersions,
Eqs. (3) and (4), leads to the appearance of singularities at
\begin{eqnarray}
\xi _1^{B,A} &=&\xi _{B,A}^{-}\left( p_z=0\right) -S, \nonumber
\\ \xi _2^{B,A} &=&\xi _{B,A}^{+}\left( p_z=0\right) -S, \nonumber
\\ \xi _3^{B,A} &=&t+\left( 1-\alpha \right)
\frac{t_{yz}^2}{8t}+\left( 2\alpha -1\right) S+\alpha \left[
t_{yz}+\sqrt{\left( t+t_{yz}\right) ^2+\left( \frac U2\right)
^2}\right] , \nonumber
\end{eqnarray}
where
\begin{eqnarray}
S &=&\frac 12\left[ \sqrt{\left( t+t_{yz}\right) ^2+\left( \frac
U2\right) ^2}-\sqrt{\left( t+\frac{t_{yz}^2}{8t}\right) ^2+\left(
\frac U2\right) ^2}+\frac{t_{yz}^2}{8t}-t_{yz}\right], \nonumber
\end{eqnarray}
and $\alpha $ takes sign $+1$ and $-1$ for the correlated subbands
$B$ (Eq. (3)) and $A$ (Eq. (4)), respectively. The van Hove square
root divergencies inside the correlated bands are manifestations
of the quasi-one-dimensional electronic structure.

The comparison of the calculated electronic structure with the
measured optical transitions allows us to estimate the
Anderson-Hubbard parameter and the hopping energies of correlated
electrons in the zigzag ladder compound SrCuO$_2$.

\section{Experimental results}
Figure 6 shows the pseudo-dielectric function
$\varepsilon(\omega)=\varepsilon_1(\omega)+i\varepsilon_2(\omega)$
for a light polarized parallel to the {\bf a}- axis (Fig. 6(a)),
and the {\bf c}-axis (Fig. 6(b)). As it can be seen from Fig.
6(a), the maximum of $\varepsilon_2(\omega)$ is at about 1.77 eV.
This value corresponds to the energy gap along the {\bf a}-axis
direction, which is perpendicular to the CuO chains. Besides the
most intensive peak at 1.77 eV, we found the next peak at about 2
eV. To determine the precise energies of these electronic
transitions we fitted both the real and the imaginary parts of the
second derivative spectra $d^2\varepsilon/dE^2$ simultaneously by
a least squares routine in terms of standard line shapes:
$$\varepsilon(\omega)=C-Ae^{i\phi}(\omega-E_0+i\Gamma)^m,$$ where
$A$, $E_0$ and $\Gamma$ are the amplitude, energy and half
linewidth of the electron transitions, respectively. The $\phi$ is
a phase factor and the exponent $m$ has the value $-1/2$ for the
one-dimensional case. The results of the calculations are given in
the inset of Fig. 6(a). At higher energies we observed two
additional peaks at about 4.1 eV and 5.1 eV. The origin of all
these transitions will be discussed later. For E$||$c polarization
the first clearly pronounced peak appears at about 3.2 eV, and the
next maxima is at about 5.2 eV (see Fig. 6(b)). For this
polarization the low intensity of the dielectric function does not
allow us to apply the same fitting procedure as in the E$||$a
case.

An unpolarized reflectivity spectrum is shown in Fig. 7(a) in the
energy range from 1 to 5.5 eV. The peak positions at 1.42 eV, 1.73
eV, 2.04 eV, 2.5 eV, 2.9 eV, 4 eV and 5.2 eV are determined as the
maxima of an optical conductivity function obtained from the
Kramers-Kronig analysis of the reflectivity spectrum. In order to
distinguish different polarization contributions in the
reflectivity spectrum shown in Fig. 7(a), we calculated the
polarized reflectivity spectra using the pseudo-dielectric
function data from Fig. 6. The results of the calculations are
given in Fig. 7(b). As it can be seen from Fig. 7(b), the main
contribution to the unpolarized spectra comes from the E$||$a
polarization. As a proof that the lowest intensity peak in
unpolarized reflectivity spectra belongs to the E$||$c
polarization, we measured polarized photoreflectance spectra in
the energy range below 1.5 eV, which are shown in the inset of
Fig. 7(a). These spectra clearly demonstrate that the reflectivity
maximum at the lowest energy in Fig. 7(a) originates from the
E$||$c polarization. The peak position of this transition is found
at 1.43 eV using a photoreflectance fitting line shape procedure
based on the third derivative of the primary spectra \cite
{Aspnes}. Thus, we conclude that {\it the lowest energy gap in
SrCuO$_2$ is at about 1.42 eV for the polarization along the Cu-O
chains.}

Figure 8 represents the absorption spectra of SrCuO$_2$ measured
at room temperature and 5K for the incident light polarized along
the {\bf c}- and the {\bf a}-axis. These spectra were calculated
using the relation $\alpha
(\omega)=(1/d)\ln[(1-R(\omega))^2/T(\omega)]$, where $R$ and $T$
represent the reflectivity and transmission coefficients, while
$d$ is the thickness of the sample. As it can be seen from the
Fig. 8, there is a strong anisotropy in the positions of the
absorption edges for these polarizations.

\section{Discussion}
The experimental results given in Figs. 6-8 can be summarized as
follows:

({\it i}) A strong anisotropy of optical properties is clearly
seen. In the direction perpendicular to the Cu-O chains (E$||$a)
the spectral weight is centered at about 1.8 eV. All other
transitions at higher energies have smaller contributions to the
dielectric function for this polarization. In the case of the
E$||$c polarization (Fig. 6(b)) the spectra show very low
intensity and the spectral weight for this polarization is shifted
to higher energies at about 3.2 eV and 5.2 eV.

({\it ii}) The lowest energy gap appears at about 1.42 eV, well
bellow the charge transfer gap (1.7 - 2.0 eV) of other CuO based
materials.

({\it iii}) The lowest energy gap for E{$||$c} shows a very small
energy change as the temperature is lowered, whereas the
absorption edge of the corresponding gap for the E$||$a shifts to
higher energies.

({\it iv}) The higher energy electron transitions appear at about
4.1 eV and 5.2 eV for both polarizations.

These features can be successfully explained in the framework of
the electronic structure calculations, Sect. III.B. The parameters
of our model for correlated electrons, $t$, $t_{yz}$ and $U$, were
fitted to the experimental value of the indirect $\Delta_0$ gap
($\Delta_0$=1.42 eV) and an exchange energy along the chain (leg)
$J=0.18$ eV \cite{Matsuda}. We used also the ratio of exchange
energies $J'/J$=0.16 ($J'$ represents the exchange energy between
legs), because this value is in the middle of the range
$J'/J=0.1-0.2$ proposed for this kind of copper oxides
\cite{Rice}. Using the relation between exchange and hopping
energies $J=4t^2/U$ and Eq. (5) for the lowest energy gap, we
estimated the other model parameters, $t$=0.30 eV, $t_{yz}$=0.12
eV, $U=2.0$ eV. The obtained magnitude of the Anderson-Hubbard
parameter of SrCuO$_2$ somewhat lower then the one ($U=2.1$ eV)
obtained previously by us in the electronic structure study of
Sr$_{14}$Cu$_{24}$O$_{41}$ \cite{Popovic1}. The hopping energy
$t$=0.3 eV along the legs is larger than in the case of
Sr$_{14}$Cu$_{24}$O$_{41}$ ($t$=0.26 eV) due to the difference
between the exchange coupling constants: $J=180$ meV in SrCuO$_2$
is larger than the $J=128$ meV value in
Sr$_{14}$Cu$_{24}$O$_{41}$. The fact of the coincidence of the
Anderson-Hubbard parameters $U$ for SrCuO$_2$ and
Sr$_{14}$Cu$_{24}$O$_{41}$ is natural to expect because the local
environment of Cu$^{2+}$ ions is similar in these oxides and the
on-site Coulomb repulsion is governed mainly by the electron
interactions on the same copper $d$-orbital.

The energy vs. wavenumber curves for correlated electrons (Fig. 5
(a)) show a large dispersion along the $k_z$ direction (parallel
to the Cu-O chains, along the {\bf c}-axis) and no dispersion
along the $k_x$ direction, in agreement with the LSDA calculations
(Fig.2). Nevertheless, the lowest energy gap at about 1.42 eV
appears for the transition from the Z to the $\Gamma$ point of the
BZ (indirect transition, denoted as $\Delta_0$ in Fig. 5 (a)). The
direct transition gaps $\Delta_2$, $\Delta_3$ and $\Delta_4$ at
$\Gamma$ (X) point of the BZ, are found to be 1.88 eV, 2.05 eV,
and 2.19 eV, respectively (see Fig. 5(b)).

According to our calculations in Sec.III.B, the correlation gap
$\Delta_0$, Eq.(5), represents the lowest energy transitions
between the split $3d_{y^2-z^2}-3d_{y^2-z^2}$ states within each
Cu atom. Thus, it is natural to expect a negligible temperature
dependence of the correlation gap. In fact, experimental data
presented in Fig. 8 show no temperature shift of the absorption
edge for the E$||$c polarization. It means that SrCuO$_2$ belongs
to the group of low-dimensional insulators with a correlation gap.
Further support for this assumption can be found in the case of
Sr$_2$CuO$_3$, also 1 D cuprate. Maiti {\it et al.} \cite {Maiti}
assigned the gap, observed at about 1.5 eV in Sr$_2$CuO$_3$, as an
insulating gap of the correlated nature. Nevertheless, our Raman
spectroscopy study \cite {Popovic} clearly demonstrated that the
resonance behavior in SrCuO$_2$ differs strongly from that in the
2D cuprates. Namely, the resonant Raman scattering shows that the
infrared active modes and their overtones resonate much more
strongly for the laser energies near the charge transfer-gap in
the 2D insulating cuprates \cite{Abrasev1}. In 1D SrCuO$_2$ the
infrared modes resonate at energies noticeably higher than the
correlation gap $\Delta_0$ \cite {Popovic}. Finally, we have found
that a similar spectral weight distribution as the one given in
Fig. 6(b) is also reported for Li$_2$CuO$_2$ for the light
polarized along the chains \cite{Mizuno}. This supports once again
our findings that the electronic structure of the 1D cuprates
differs significantly from that of the 2D cuprates.

In the inset of the Fig. 5(a) we show the $\Delta_0$ gap
dependence on the $U$ parameter, Eq.(5), for the two different
values of the interchain hopping $t_{yz}$ at $t$=0.30 eV. The
limiting case $t_{yz}$=0 represents one leg of the ladder or a
single-chain structure which exists in Sr$_2$CuO$_3$. As it is
seen in the inset of Fig. 5(a), the interchain hopping reduces a
gap. It means that the corresponding gaps in the single-chain
compound Sr$_2$CuO$_3$ should be somewhat larger than in
SrCuO$_2$. This agrees well with our experimental findings.
Namely, the absorption edge in Sr$_2$CuO$_3$ is at 1.5 eV \cite
{Maiti}, about 5{\%} higher than in SrCuO$_2$, see Fig.8.

The electron structure calculations for correlated electrons (Fig.
5) make it possible to compare to experimental data not only the
lowest energy gap value. The peaks at 1.77 eV and 2 eV of the E$||
$a $\varepsilon_2(\omega)$ spectra correspond to the
L$_{2-}$-L$_{1+}$ and L$_{2-}$-L$_{2+}$ transitions at $\Gamma$
and X-points ($\Delta_2$ and $\Delta_3$ gaps), respectively. In
the E$|| $c polarized $\varepsilon_2(\omega)$ spectra we observed
three very low intensity peaks at about 2.2 eV, 2.6 eV and 3 eV.
These transitions can be assigned as L$_{1-}-$L$_{1+}$
($\Delta_4$), L$_{2-}-$L$_{3+}$ and L$_{1-}-$L$_{3+}$,
respectively. Because of the very low intensity of these modes, it
was hardly possible to extract their exact positions from the
noise level. The energies of all possible electron transitions
between occupied and empty states are shown in the inset of Fig.
5(b).

Now we discuss the two higher energy features at about 4.1 eV and
5.2 eV. They have been already observed in a lot of cuprates, as
discussed in Ref. \cite {Alonso}. Alonso {\it et al.} \cite
{Alonso} shown that in Nd$_{2-x}$Ce$_x$CuO$_4$ the transition at
about 4 eV originates from the Cu $3d$ - Cu $4p$ intraionic
transition and the highest energy transition about 5.2 eV occurs
due to transition between the O $2p$ and the Cu $4s$ states. The
existing band structure calculations of SrCuO$_2$ are not detailed
enough to corroborate this assignment.

Let us now consider Fig. 8, where the absorption coefficient
spectra are presented. There is a large difference between the
absorption edge positions for the E$|| $c and the E$|| $a
polarizations. Besides that, the slope of the absorption
coefficient is higher for the E$|| $c then the E$|| $a
polarization, as a consequence of a quasi 1D character of this
transition. A general approach to analyze the fundamental
absorption edge is based on the use of the power-law behavior of
$\alpha(\omega)$ in the vicinity of the band gap, $E_{gap}$. The
absorption coefficient can be described as $\alpha (\omega)=
A(\omega-E_{gap})^{k}$, where $A$ is a slowly varying function
which is regarded as a constant over the narrow range under study,
and a number $k$ depends on the nature of an electron transition
from the occupied to the empty band. The $k$ value is 1/2 for a
direct transition and 2 for an indirect one. Thus, the
$\sqrt{\alpha(\omega)}\sim(\omega-E_{gap})$ dependence can be used
to obtain the indirect gap value by extrapolating the linear
portion of this curve to intersect with the $\omega$ axis. The
direct gap position can be extracted from a maximum of the first
derivative of $\alpha(\omega)$
($\alpha'(\omega)\sim1/\sqrt{\omega-E_{gap}}$). This procedure is
valid for a three dimensional case and for the parabolic electron
energy dispersions.

The absorption spectra for strongly correlated electron systems
were analyzed using the density of states vs. energy dependence of
1D \cite {Dressel} or 3D \cite{Zibold} semiconductors with
parabolic energy dispersions. Actually, the dispersion branches in
strongly correlated systems are parabolic in the vicinity of the
high symmetry points of the BZ: cos $p= 1+p^2/2$ around $\Gamma$
($\simeq0$) and cos $p=1-p^2/2$ around Z ($\simeq\pi$).
Furthermore, from the extrapolation of the linear part of the
$\alpha(\omega)$ to the energy axis, we obtain the value of 1.52
eV (1.65 eV) for E$||$c (E$||$a) polarization. This energy is
higher (lower), than that obtained by ellipsometric or
reflectivity measurements. The complete agreement with
ellipsometric and reflectivity data is achieved when we consider
{\it the E$||$c absorption edge as an indirect transition and the
E$||$a as a direct one}, which is illustrated in the left and the
right insets of Fig. 8. Thus, we have concluded that {\it the
correlation gap in SrCuO$_2$ represents the indirect transition of
carriers between the occupied and the empty correlated subbands}.
The indirect transition requires a change in both energy and
momentum of carriers, as our band structure calculations predict,
see Fig. 5(a). It is well known that in semiconductors an electron
momentum is conserved via an interaction with phonons \cite{Yu}.
For such conclusion here further experiments are necessary to
clarify this point. They can be done by the transmission
measurements through the samples of different thickness and/or
comparison of photoluminescence with photoreflectance spectra.

Quite recently, based on the Bethe ansatz solution for the Hubbard
chain it was shown \cite {Carmelo}, that in an insulating state
the optical conductivity can be described as $\sigma(\omega)\sim
C(\omega-E_{MH})^{1/2}$, where $E_{MH}$ is the (Mott-Hubbard) gap.
Since $\alpha(\omega)\sim \sigma(\omega) $, this dependence has
the same power-law as the one we used for the direct electron
transition.

For the E$||$a polarization, the absorption edge shifts to higher
energies by about 0.06 eV, when the temperature decreases from a
room temperature to 5 K. For this electronic transition we cannot
neglect hybridization of the Cu $3d$ with the O $2p$ bands. Thus,
the absorption edge shifts to higher energies according to the
change of the Cu-O distance with temperature. Besides that, this
energy gap represents the transition between the dispersionless
branches with very high effective electron masses and,
consequently, with a strong influence of ligands. Because of that
we believe that for the E$||$a polarization the absorption edge
shift appears mainly due to dilatation of lattice by lowering the
temperature.

In conclusion, we have investigated the electronic structure of
the zigzag chain compound SrCuO$_2$ combining polarized optical
absorption, reflection, photoreflectance and pseudo-dielectric
function measurements with theoretical estimations. At 300 K these
measurements yield the energy gaps 1.42 eV and 1.77 eV along and
perpendicular to the Cu-O chains, respectively. The electronic
structure of this oxide is calculated using the LSDA with gradient
corrections and the tight-binding method for the correlated
electrons. The gap value of 1.42 eV (1.86 eV) is found for the
electron hopping energies between copper sites along legs,
$t=0.30$ eV, and between them, $t_{yz}$= 0.12 eV, with the
Anderson-Hubbard parameter $U=2.0$ eV. The obtained experimental
results and electronic structure calculations have shown that the
SrCuO$_2$ zigzag chain compound belongs to the low-dimensional
insulators with the band gap of the correlated nature.

\section*{Acknowledgments}

Z. V. P., V. A. I. and O. P. K. acknowledge support from the
Research Council of the K. U. Leuven and DWTC. The work at the K.
U. Leuven is supported by the Belgian IUAP and Flemish FWO and GOA
Programs. M. J. K. and Z. V. P. thank Roman Herzog - AvH - Bonn
and University of Valencia, respectively, for partial financial
support. V. A. I. acknowledges Yu. M. Kagan for fruitful
discussions.

\clearpage

\begin{figure}
\caption {Crystal structure of SrCuO$_2$ in the (a) (100) and (b)
(001) plane. Dashed line rectangle represents a unit cell of the
zigzag ladder.} \label{fig1}
\end{figure}

\begin{figure}
\caption {Electronic structure of SrCuO$_2$ along the several high
symmetry directions in BZ calculated using the LSDA with gradient
corrections method. The cartesian coordinates of high-symmetry
points are as follows: $\Gamma$ (0,0,0); Y(2$\pi/a$,0,0),
(0,2$\pi/b$,0); Z(0,0,$\pi/c$); T(2$\pi/a$,0,$\pi/c$),
(0,2$\pi/b$,$\pi/c$). } \label{fig2}
\end{figure}

\begin{figure}
\caption {Total density of states calculated within the LSDA with
gradient corrections method. Inset:  Partial density of states
close to Fermi level of different atoms of SrCuO$_2$.}
\label{fig3}
\end{figure}

\begin{figure}
\caption{The contribution of different orbitals to the calculated
partial density of states (DOS) of SrCuO$_2$. (a) Cu-$d$ DOS, (b)
O1-$p$ DOS, and (c) O2-$p$ DOS.} \label{fig4}
\end{figure}

\begin{figure}
\caption {(a) The tight-binding dispersions for correlated
 electrons in SrCuO$_{2}$ with parameters $t$=0.3 eV,
$t_{yz}$=0.12 eV, U=2.0 eV. The momenta are given in units $\mid
p_y\sqrt{2}\mid$=$\mid p_z\mid $=$\pi$ of the Brillouin zone
boundaries, the Fermi energy $E_F=0$ is inside of the correlation
gap. Inset: energy gap vs. $U$ dependence for $t_{yz}=0$ and 0.12
eV; (b) The correlated electron density of states as a function of
energy. The electronic structure parameters are the same as for
Fig.5(a). Inset: the energy values of all possible transitions
between occupied and empty states.} \label{fig5}
\end{figure}

\begin{figure}
\caption {Room temperature real ($\varepsilon_1$) and imaginary
($\varepsilon_2$) part of the pseudodielectric function of
SrCuO$_2$. The spectra of the (010) surface taken with (a) the
a-axis (E$|| $a) and (b) the c-axis (E$|| $c), parallel to the
plane of incidence. Inset: second derivative of dielectric
functions for E$|| $a polarization in the 1.5 - 2.5 eV spectral
range.} \label{fig6}
\end{figure}

\begin{figure}
\caption {(a) Room temperature unpolarized reflectivity spectra of
the SrCuO$_2$ single crystal. (b) Reflectivity spectra calculated
using experimental data of the pseudo-dielectric function from
Fig. 6. Inset: polarized photoreflectance spectra at room
temperature.} \label{fig7}
\end{figure}

\begin{figure}
\caption {Absorption coefficient spectra of SrCuO$_2$ at room and
liquid helium temperatures. Inset, left panel:
$\sqrt{\alpha(\omega)}$ vs. $\omega$ dependence. Inset, right
panel: first derivative of $\alpha(\omega)$ dependence for E$|| $a
polarization. } \label{fig8}
\end{figure}

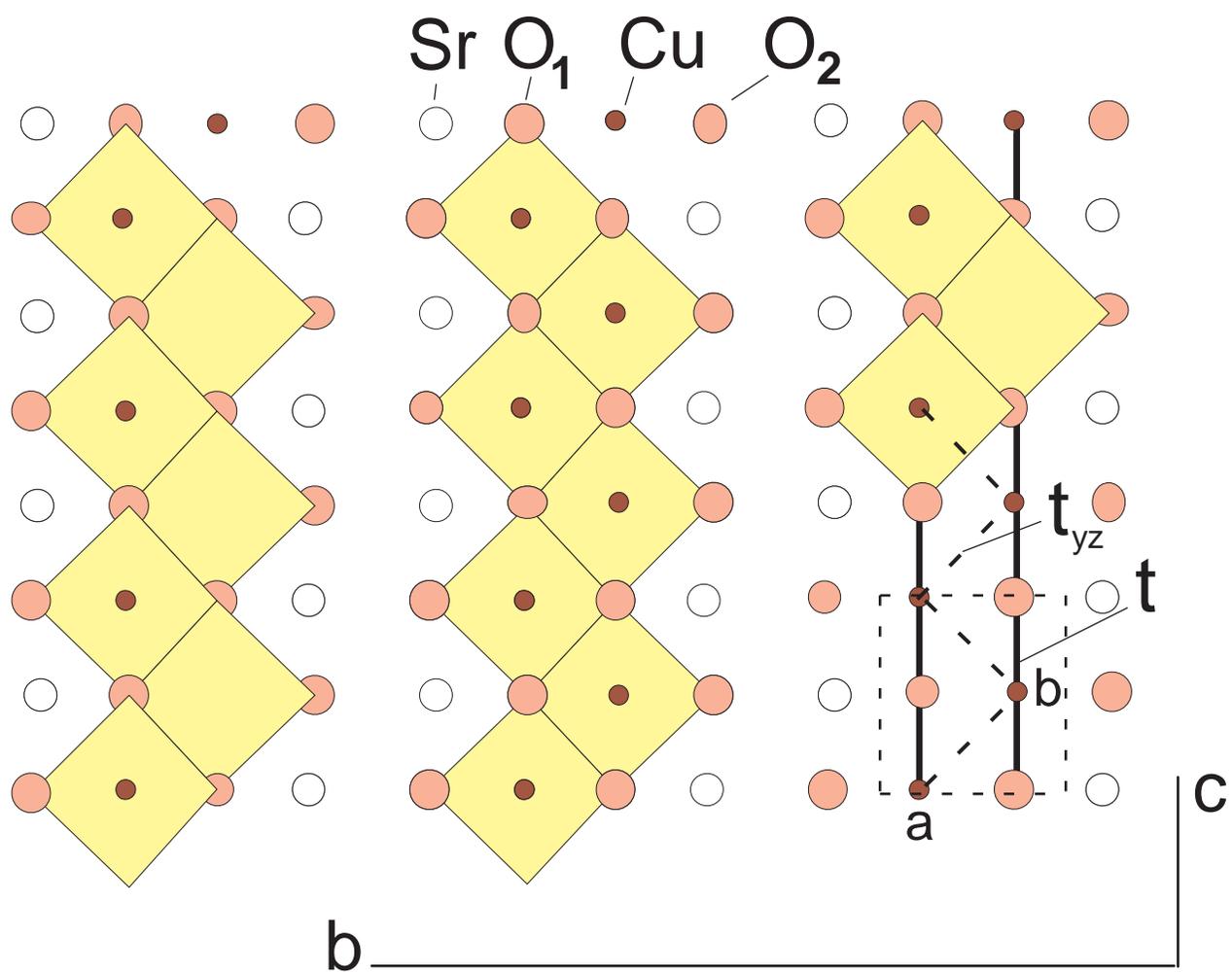
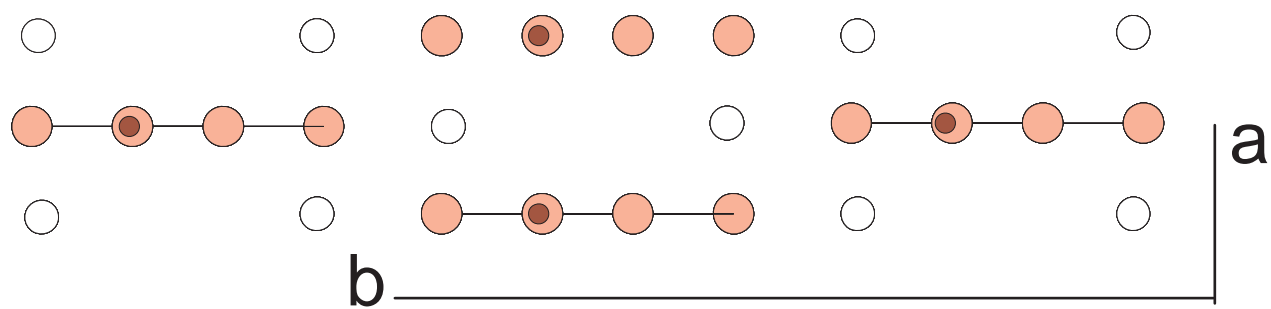
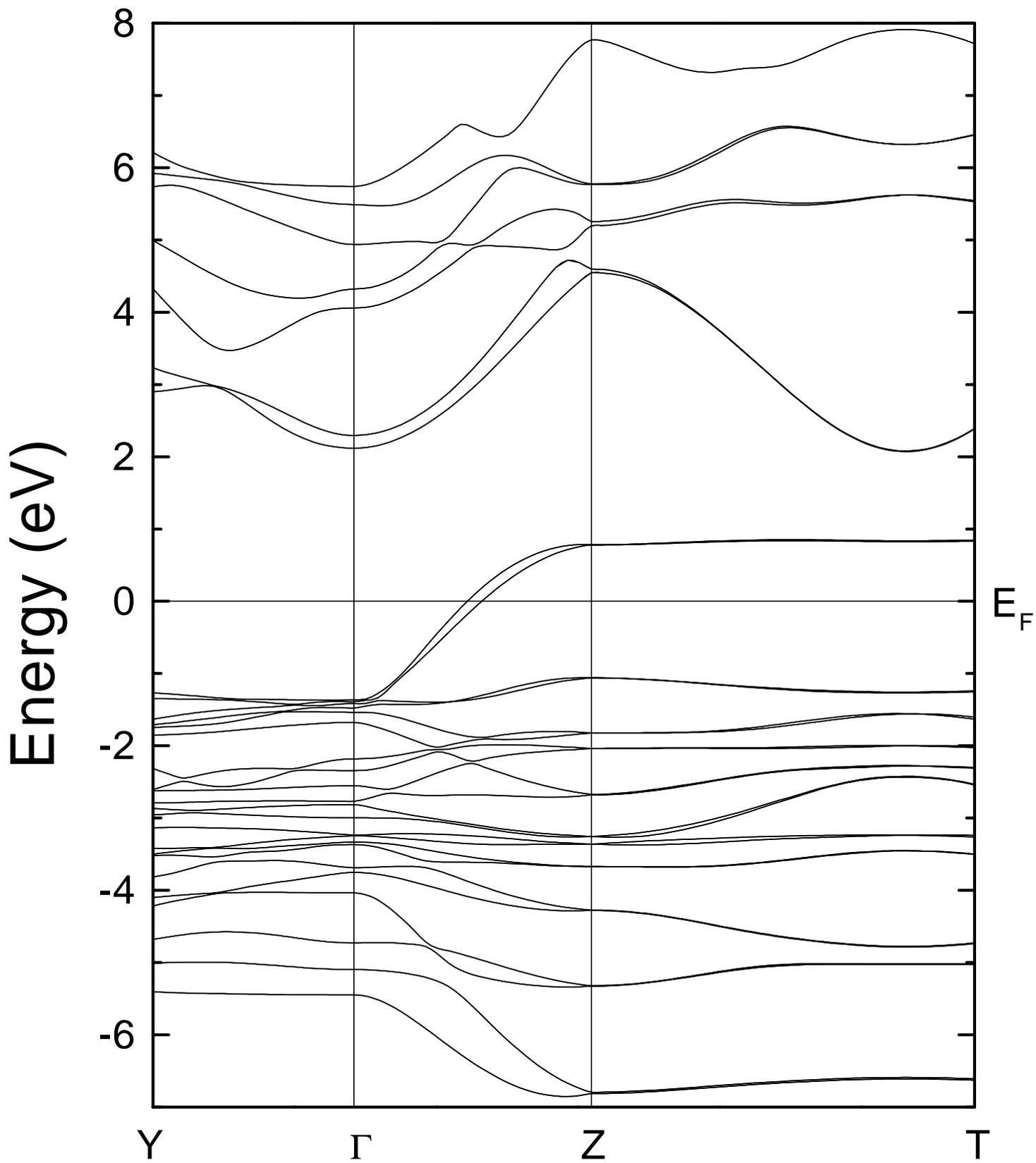
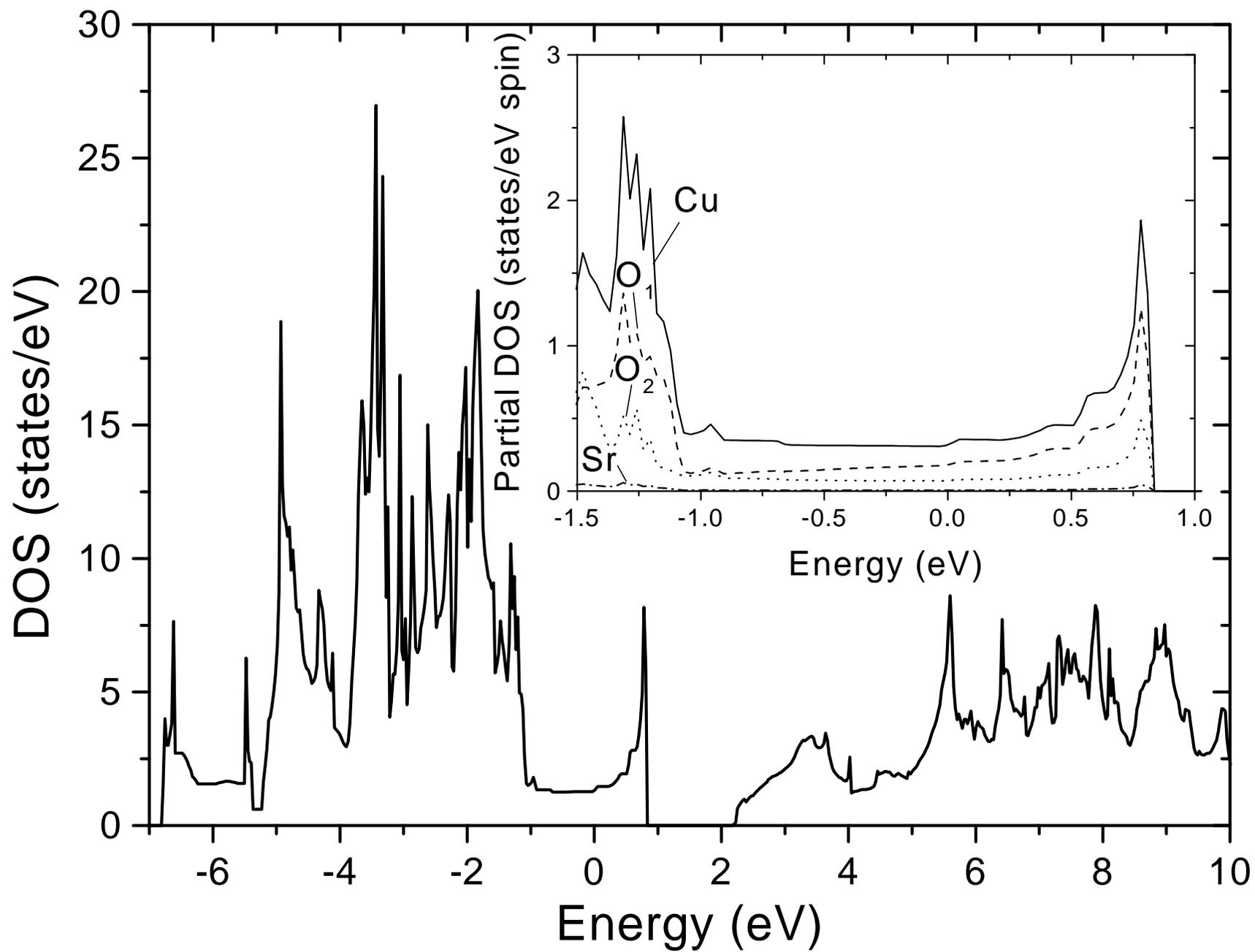
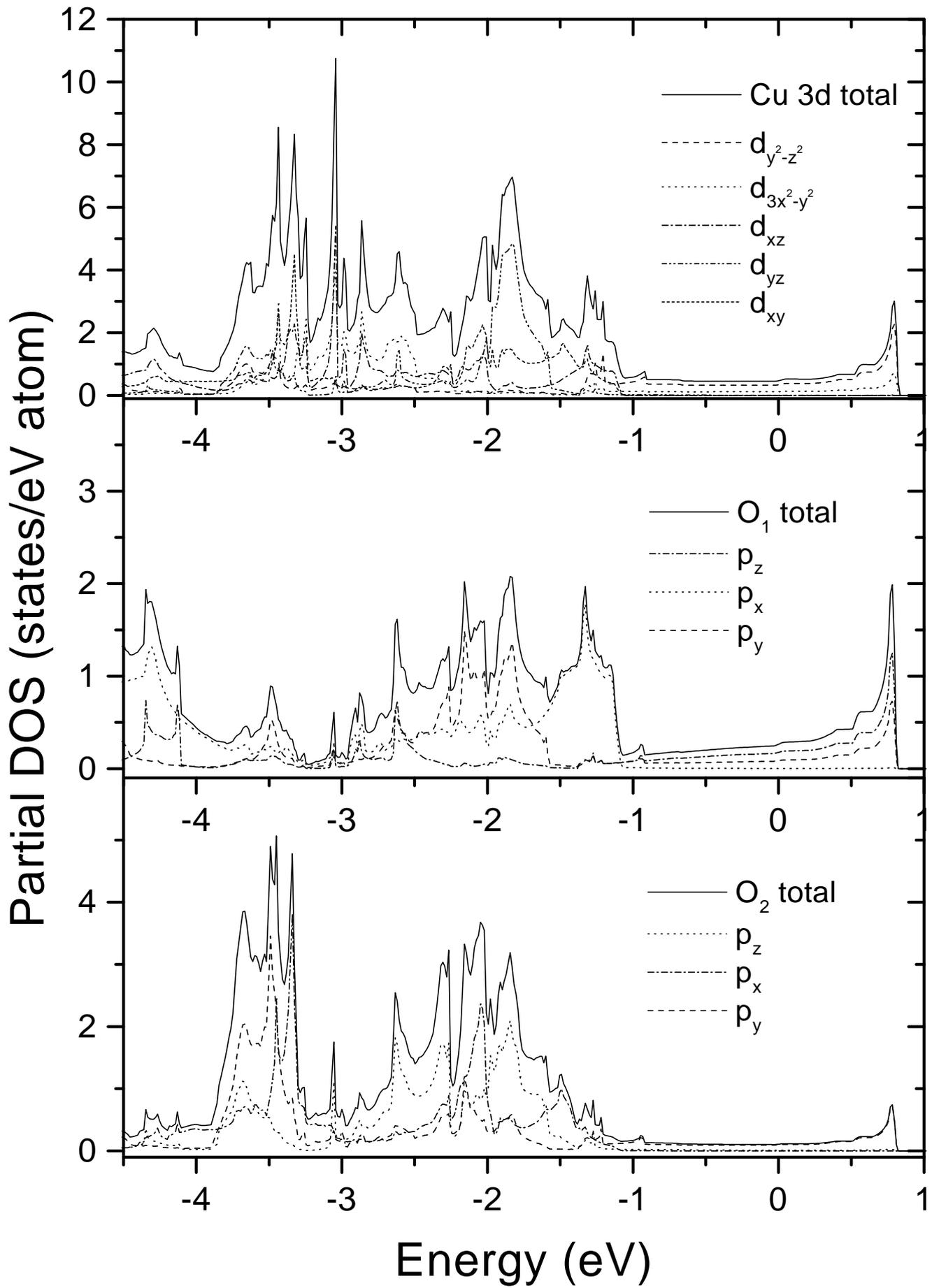
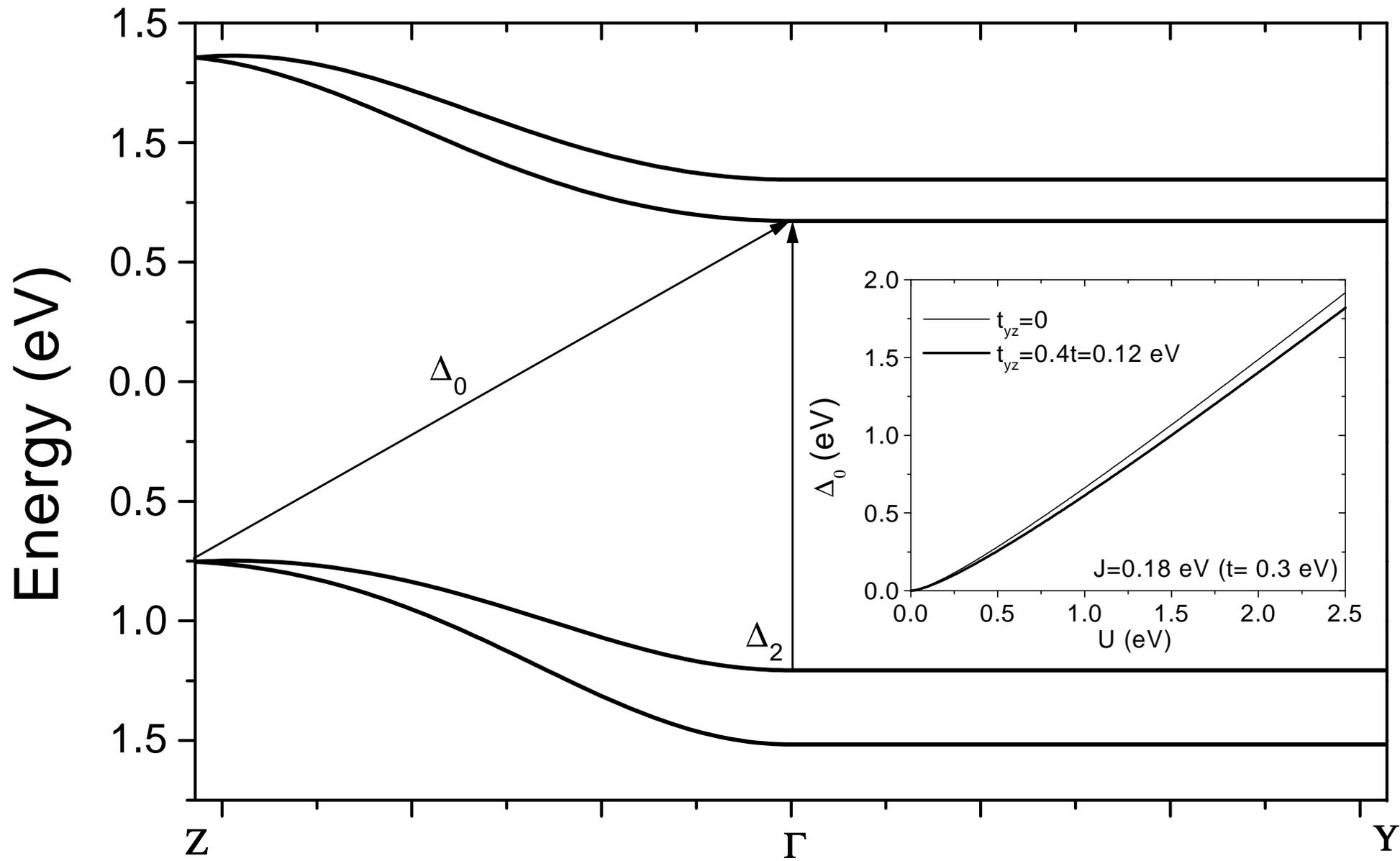
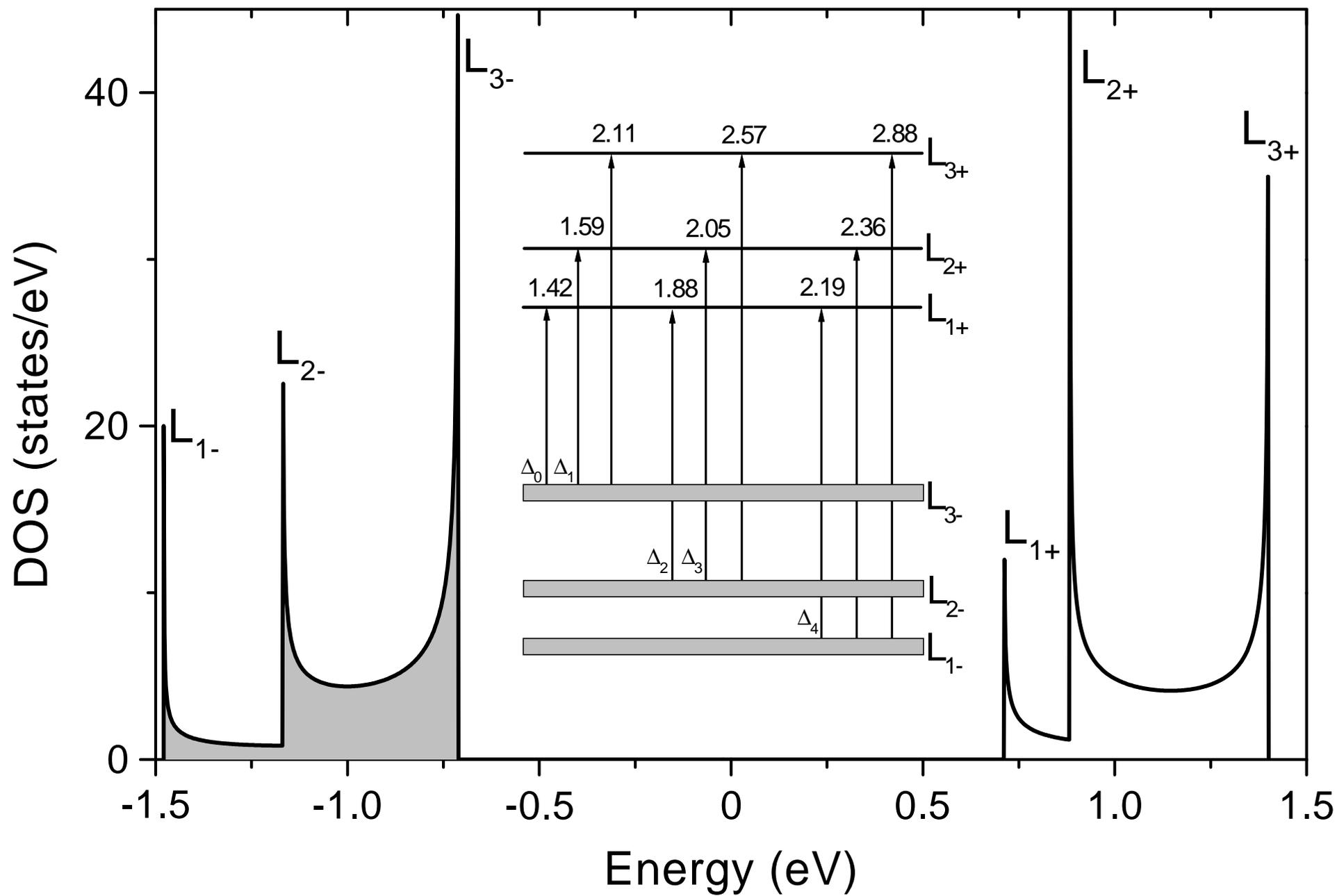
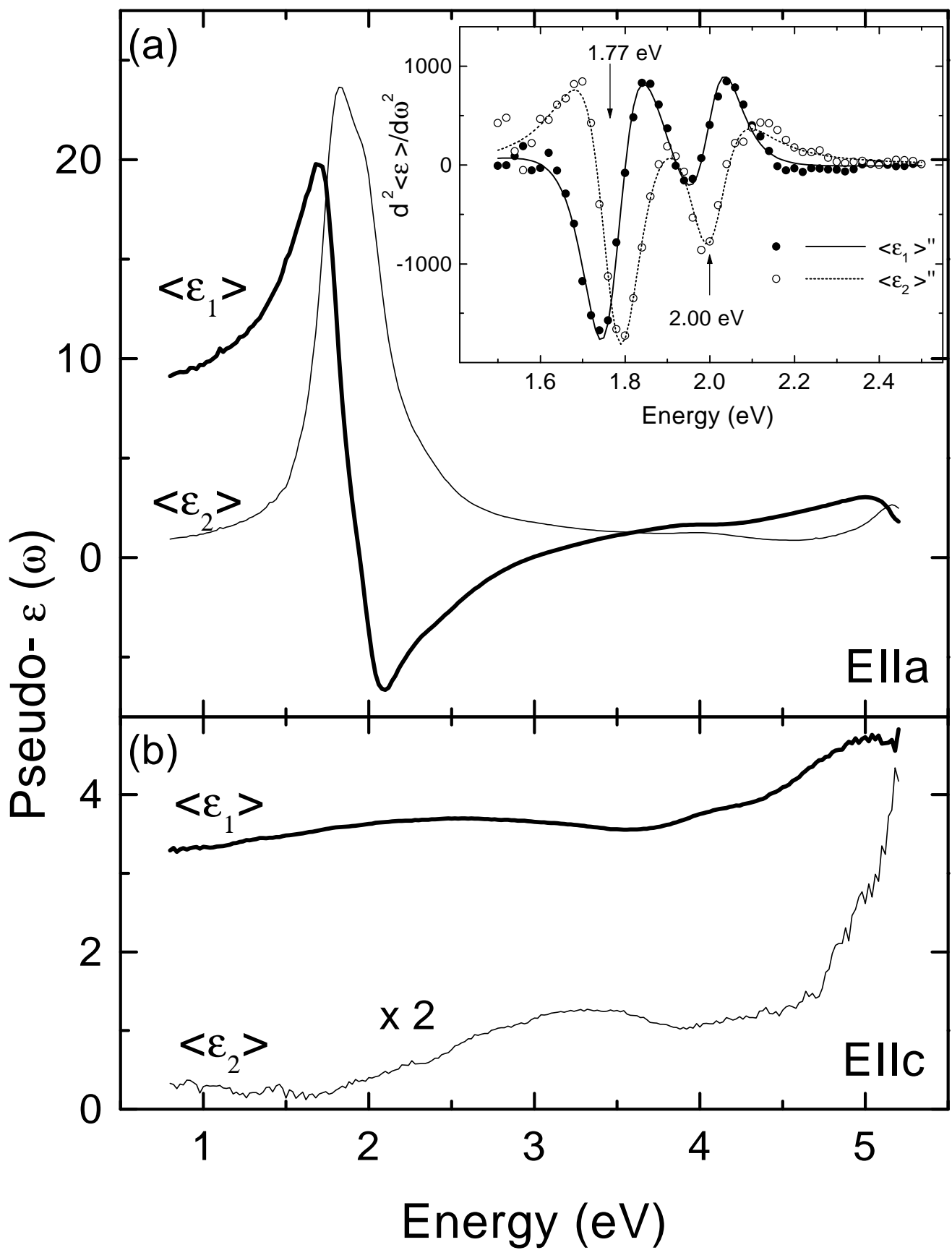
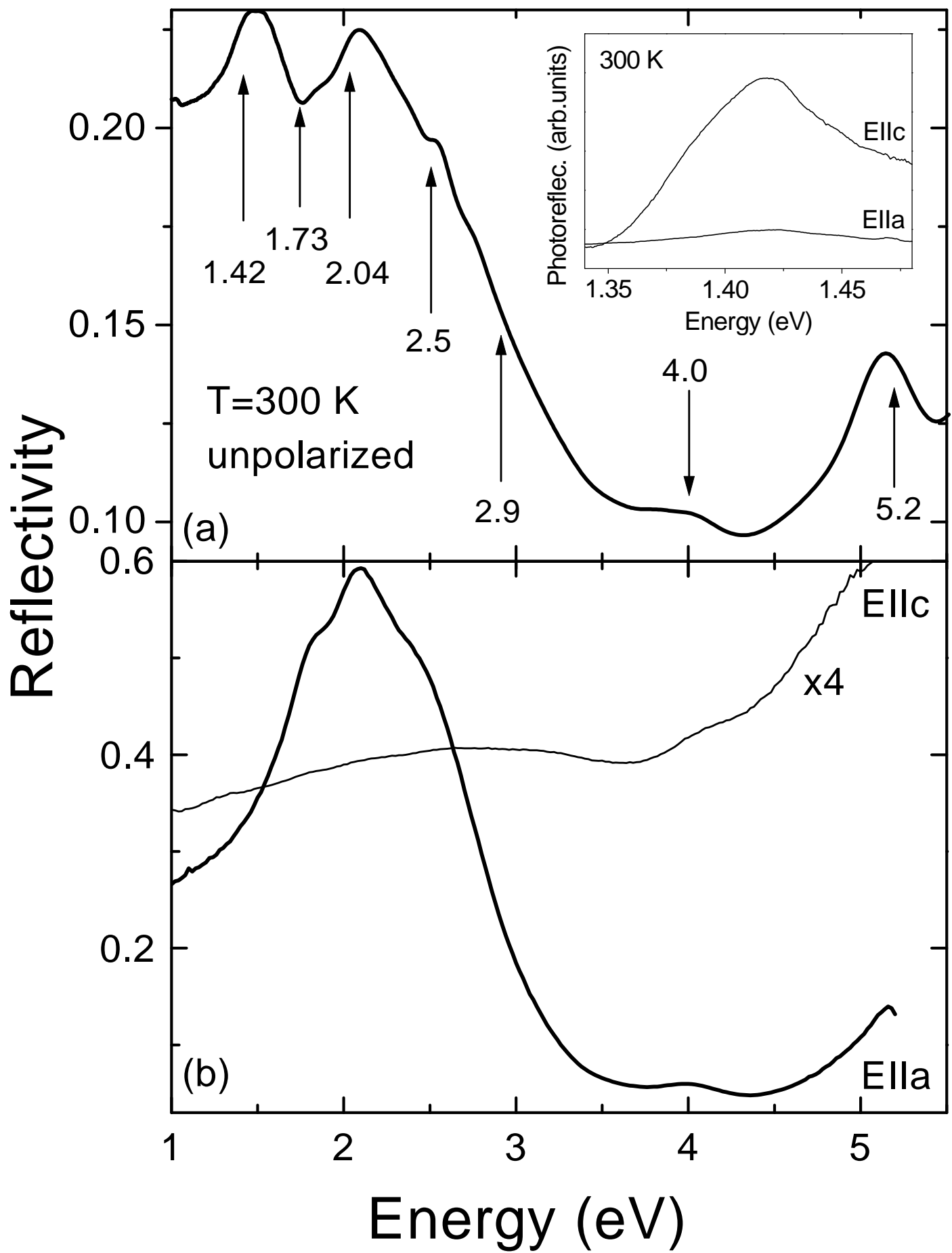
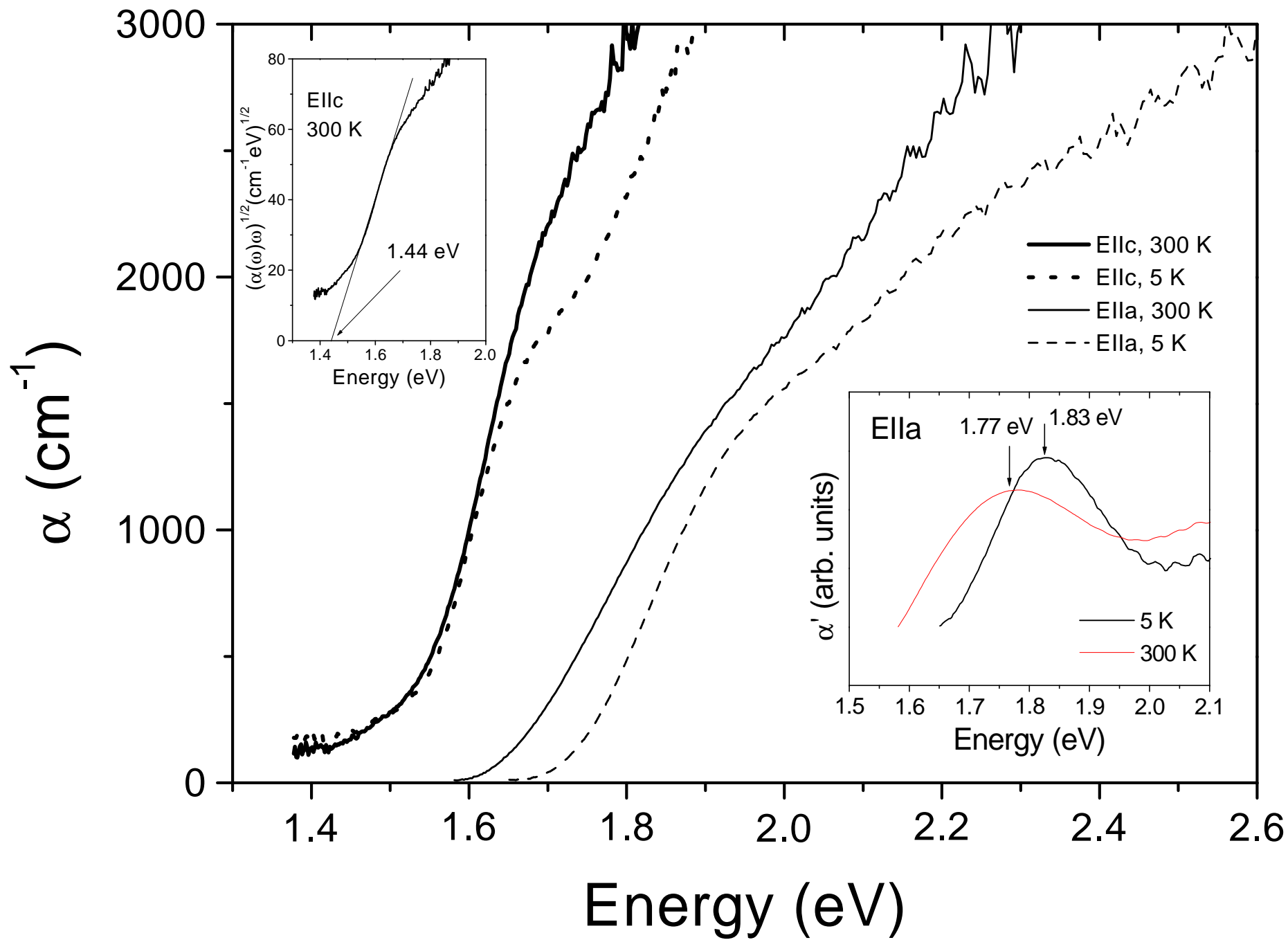
\end{document}